# Nuclear Thermo-Electric Thruster

Christianna Jessica Tymczak[1]

[1)] *Houston Community College, Houston, Texas*
*christianna.tymczak@hccs.edu*

We present a theoretical analysis of an innovative combination of a nuclear thermal and electromagnetic (EM) thruster. Specifically, we scrutinize the thermodynamics involved in integrating a nuclear thermal reactor with an expansion turbine. This configuration facilitates the generation of substantial electrical power, which is then utilized to power an EM thruster (similar to an afterburner). This process results in a notable increase in the ISP from 900 to 1200 without the necessity for thermal radiators. Furthermore, by incorporating thermal radiators, the ISP can be further increased to approximately 4000. This enhancement allows for a significant reduction in transit time to destinations such as Mars and the outer and inner planets. We provide several examples to illustrate the potential applications of this innovative propulsion system.

## I. Introduction

Rapid transit to the inner and outer planets is essential if humanity is to explore and utilize the resources of the solar system. However, significant challenges on long-duration missions will severely limit the capabilities of human exploration. The risks associated with extended missions include radiation exposure, isolation, weightlessness, and logistical and supply issues [1-4]. To address these challenges, we propose examining a hybrid nuclear thermal electric rocket engine (NTE thruster) designed to harness the virtually limitless thermal energy generated by nuclear reactors [5, 6]. This engine features adjustable specific impulse (ISP) and thrust, ranging from low ISP (1200 sec) with high thrust to high ISP (>4000) with low thrust. Figure 1 illustrates the proposed

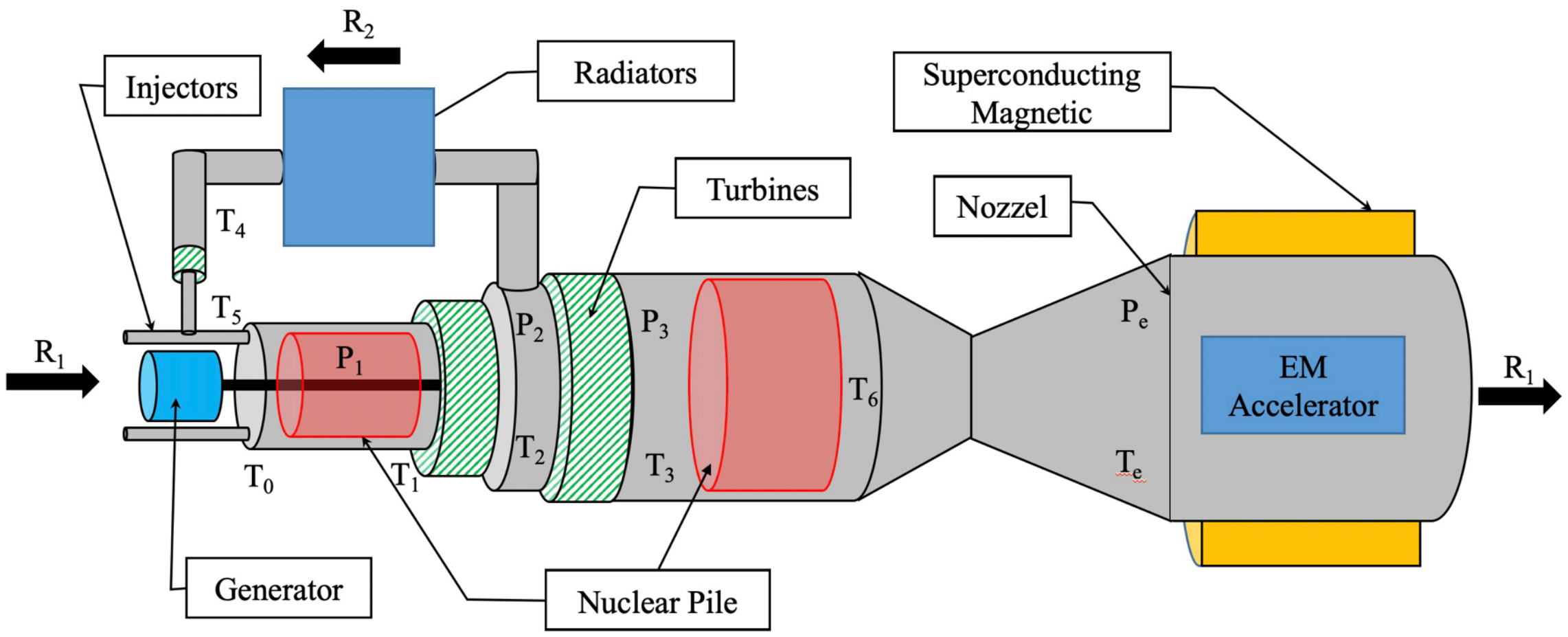

FIG 1: The Nuclear Thermal Electric Rocket Engine schematic

configuration of our NTE thruster. This engine consists of five primary components:

1) A nuclear pre-heater, where the input gas (hydrogen) undergoes high-temperature heating at elevated pressure.
2) A high-temperature gas turbine and electric generator responsible for generating electrical energy to power the electromagnetic (EM) accelerator.
3) A nuclear afterburner designed to reheat the gas back up to its operational temperature.
4) A recycle loop incorporating a thermal radiator to cool and recycle a portion of the working gas.
5) An EM accelerator, used to enhance the exhaust velocity of the working gas [7-10].

In Figure 2 we depict the thermodynamic cycles of our proposed engine [5, 11].

## II. The Brayton Cycle Thermodynamic Efficiency

Let's initiate the efficiency assessment of our engine using the Brayton Cycle [12]. The power is directly proportional to the mass flow rate multiplied by the work. Two cycles merit consideration: the portion of gas exiting the engine, denoted as $R_1$, and the proportion of gas recycled through the thermal radiators, denoted as $R_2$. If we define the total flow as $R$, we can express this as follows,

$$\begin{aligned} R_1 &= (1-\alpha)R \\ R_2 &= \alpha R \end{aligned} \tag{1}$$

Where $\alpha$ is the mixing parameter, $0 \leq \alpha \leq 1$. We consider the three power cycles separately. For the *thrust* cycle ($T_0 \rightarrow T_6 \rightarrow T_E \rightarrow T_0$), the power output is;

$$\begin{aligned} \boldsymbol{P}_1 &= R_1 W_3 \\ &= R_1\left(Q_{in,3} + Q_{in,4} - Q_E\right) \end{aligned} \tag{2}$$

For the *open power* cycle ($T_0 \rightarrow T_1 \rightarrow T_3 \rightarrow T_0$), it is;

$$\begin{aligned} \boldsymbol{P}_2 &= R_2(W_1 + W_2) \\ &= R_1\begin{pmatrix} Q_{in,1} + Q_{in,2} \\ -Q_{out,1} - Q_{out,2} \end{pmatrix} \end{aligned} \tag{3}$$

And for the *closed power* cycle with a thermal radiator ($T_5 \rightarrow T_1 \rightarrow T_2 \rightarrow T_4$), it is;

$$\boldsymbol{P}_3 = R_2 W_2 = R_2\left(Q_{in,2} - Q_{out,2}\right) \tag{4}$$

We can rewrite these as

$$\boldsymbol{P}_1 = R_1\left(\int_{T_o}^{T_6} C_p(T)dT - \int_{T_0}^{T_E} C_p(T)dT\right) \tag{5}$$

$$\boldsymbol{P}_2 = R_1\left(\int_{T_0}^{T_1} C_p(T)dT - \int_{T_0}^{T_3} C_p(T)dT\right) \tag{6}$$

$$\boldsymbol{P}_3 = R_2\left(\int_{T_5}^{T_1} C_p(T)dT - \int_{T_4}^{T_2} C_p(T)dT\right) \tag{7}$$

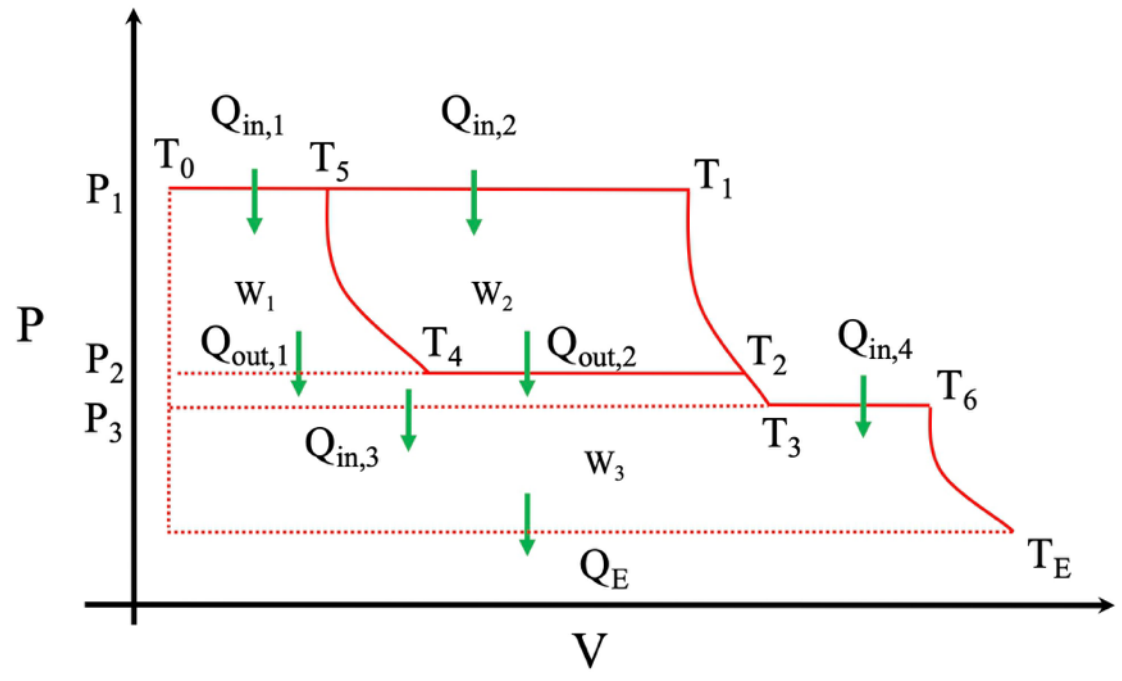

FIG 2: The thermodynamic cycles of the NTE thruster

Where the specific heat, as a function of temperature, is a known tabulated function [13], $C_P(T)$. $T_0$, $T_1$, and $T_6$ are known input parameters; and $T_2$, $T_3$, $T_4$, $T_5$, and $T_E$ must be calculated through two methods described below. Since we know the relevant pressures, we can use the adiabatic approximation for an ideal gas to compute the final temperatures [12]. However, because the specific heats vary with temperature, the adiabatic exponent will also vary with temperature and must be accounted for. We can deal with this via the infinitesimal pressure approximation (for which the exponent is approximately constant) which we detail below. For example, let's consider $T_2$; we can compute $T_2$ via the adiabatic approximation for infinitesimal pressure change through the iterative equation;

$$T[i+1] = T[i]\left(1+\frac{\delta P}{P[i]}\right)^{\left(\frac{\gamma(T[i])-1}{\gamma(T[i])}\right)} \quad (8)$$
$$i \rightarrow 1 \; to \; N-1$$

Where

$$T[1] = T_1$$
$$T[N] = T_2 \quad (10)$$
$$\delta P = \frac{P_2 - P_1}{N}$$

We use this method to compute both $T_2$, $T_3$, $T_5$, and $T_E$. As for $T_4$, this requires we compute the radiative losses through the thermal radiators, which we detail in the next section.

## III. Radiative Power: Thermal Radiators

Let us start with the Stefan-Boltzmann law [12], the thermal power radiated by a black body is,

$$\boldsymbol{P} = \sigma \epsilon A T^4 \quad (11)$$

Where $T$ is the temperature in kelvin, $\sigma$ is the Stefan-Boltzmann constant, $A$ is the area of the radiator, and $\epsilon$ is the emissivity. We will consider emissivity to be one for simplicity. We can rewrite the above equation as;

$$dQ = \sigma\,(w\,l)\,T^4 dt = C_p(T)\,m\,dT \quad (12)$$

Where $m$ is the mass and $C_P(T)$ is the specific heat. Rearranging, and using the definition of the density, we get

$$C_p(T)\frac{dT}{T^4} = \frac{\sigma}{\rho\,h}dt \quad (13)$$

Which gives

$$\int_{T_i}^{T_f} C_p(T)\frac{dT}{T^4} = \frac{\sigma}{\rho\,h}\Delta t \quad (14)$$

Where $T_i$ is the initial temperature (known) and $T_f$ is the final temperature. Using,

$$dm = \rho\,w\,h\,dl = R\,dt \quad (15)$$

And

$$v = \frac{R}{\rho\,w\,h} = \frac{l}{\Delta t} \quad (16)$$

we finally obtain,

$$\int_{T_i}^{T_f} C_p(T)\frac{dT}{T^4} = \frac{\sigma A}{R} \quad (17)$$

which we can use to calculate $T_f$

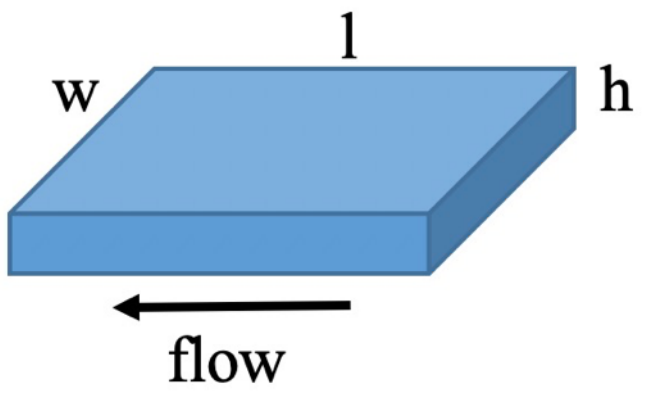

FIG 3: Thermal radiator panel

## IV. Calculation of the ISP and Thrust

Here we will assume for simplicity that 100% of the power generated is used to accelerate the propellant gas, thereby increasing the exhaust velocity. From conservation of energy, we get,

$$\frac{1}{2}\frac{\Delta N}{\Delta t} m_A v_{ex}^2 = \boldsymbol{P} \quad (18)$$

Where the rate can be related to the particle flow via,

$$\frac{\Delta N}{\Delta t} m_A = R_1 \quad (19)$$

Which, with a little algebra, gives us for the exhaust velocity, ISP, and thrust as:

$$v_{ex} = \sqrt{\frac{2\boldsymbol{P}}{R_1}}$$
$$ISP = \frac{v_{ex}}{gR_1} \quad (20)$$
$$\boldsymbol{T} = R_1 v_{ex}$$

Where

$$\boldsymbol{P} = \boldsymbol{P}_1 + \boldsymbol{P}_2 + \boldsymbol{P}_3 \quad (21)$$

## V. Results and Discussion

**Hydrogen:** For our initial example we consider Hydrogen as the propulsive gas. The initial conditions for this example are as follows:

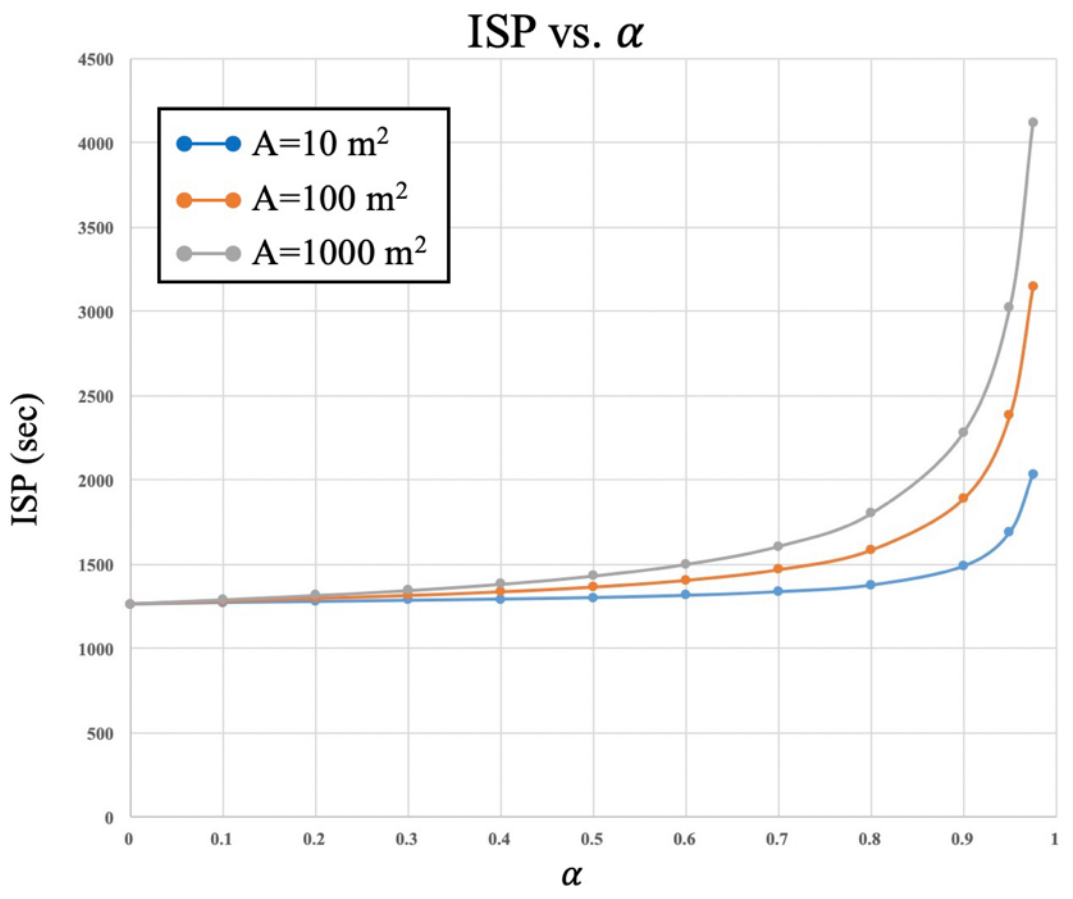

FIG 4: ISP vs. α for different thermal radiators area for Hydrogen.

$$P_1 = 300bar \qquad P_E = 0.01bar$$
$$R = 1kg/s \qquad T_0 = 100\ K$$
$$T_1 = 3000\ K \qquad T_5 = 3000\ K$$

These parameters will be kept fixed throughout the calculation. We will vary $P_2, P_3,\ A$ and $\alpha$, where $P_2$ and $P_3$ are optimized to give the highest ISP. Tables 1-3 are our tabulated results. Figure 4 shows a plot of the ISP verse the mixing parameter, $\alpha$, for Hydrogen as the propulsive gas. As can be seen, with increasing thermal surface area and an increasing mixing parameter, the ISP increases substantially. Hydrogen is by far the best choice for a rapid transit to the outer and inner planets, however several technical issues exist with this choice; i) hydrogen is notoriously difficult to store; and, ii) hydrogen issue with metal embrittlement which would need to be addressed [14].

**Helium:** For our next example we consider helium as the propulsive gas, with the initial conditions again as follows:

$$P_1 = 300bar \qquad P_E = 0.01bar$$
$$R = 1kg/s \qquad T_0 = 100\ K$$
$$T_1 = 3000\ K \qquad T_5 = 3000\ K$$

As before, we will vary $P_2, P_3,\ A$ and $\alpha$, where $P_2$ and $P_3$ are optimized to give the highest ISP. Tables 4-6 are our tabulated results. Figure 5 shows a plot of the ISP verse the mixing parameter, $\alpha$, for Helium as the propulsive gas. As can be seen, with increasing thermal surface area and an increasing mixing parameter, the ISP increases substantially. Helium has poorer performance then Hydrogen but has better thermal and chemical properties. However, its lack of availability for the inner planets would be an issue.

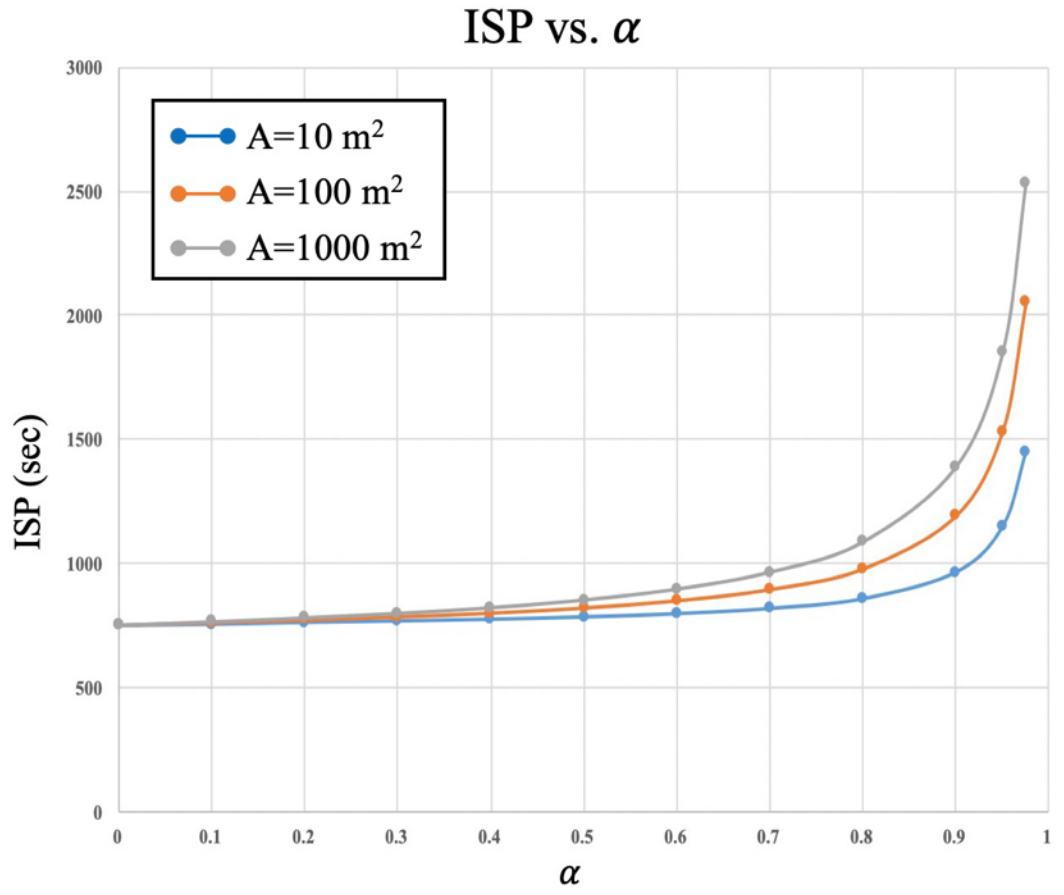

FIG 5: ISP vs. $\alpha$ for Helium and different thermal radiators area

**Water:** For our final example we consider water as the propulsive gas, the initial conditions are as follows;

$$P_1 = 300bar \qquad P_E = 0.01bar$$
$$R = 1kg/s \qquad T_0 = 100\ K$$
$$T_1 = 3000\ K \qquad T_5 = 3000\ K$$

These will be kept fixed throughout the calculation. One small note however is we need to take into account the possibility of a phase change for water via its latent heat of fusion. Tables 7-9 are our tabulated results.

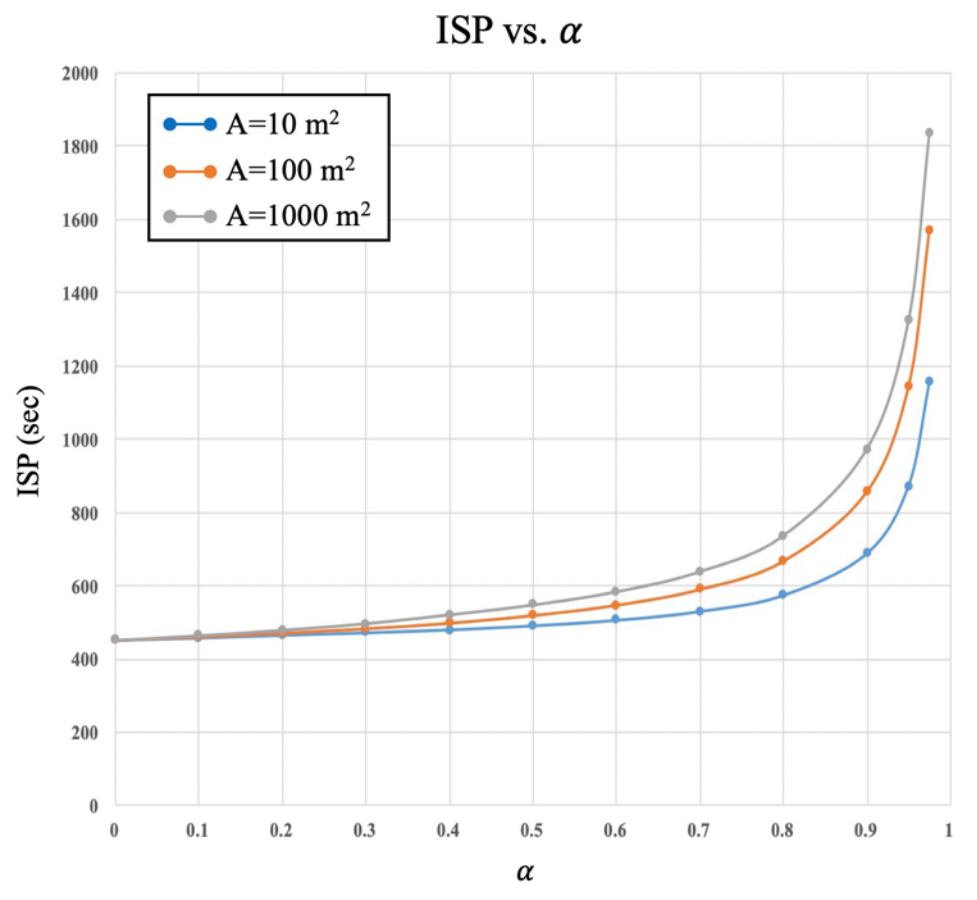

FIG 6: ISP vs. α for different thermal radiators area for Water

Figure 6 shows a plot of the ISP verse the mixing parameter, $\alpha$, for Water as the propulsive gas. As can be seen, with increasing thermal surface area and an increasing mixing parameter, the ISP increases substantially. Even though waters performance is substantially less then Hydrogen it would have significant technical advantages as a propulsive gas: i) storage of water as a propellant would be essential non-existent; ii) what has minimal issues with metal embrittlement; and iii) water is fairly available in both the outer and inner solar system (as well as ammonia, which has similar performance).

## VI. Hypothetical Mars Mission

Let us consider the orbital transfer to mars on a hypothetical hyperbolic orbit. At Earth's orbit we give our rocket an initial velocity, denoted as $\Delta \boldsymbol{v_E}$; and at the Mars orbit we give it a final velocity, denoted as $\Delta \boldsymbol{v_M}$. We consider the change in velocities to happen at a much shorter time scale then the time of flight. The initial condition at Earth are:

$$\begin{aligned} \vec{r}_i &= R_E \hat{r} \\ \overrightarrow{\Delta v_E} &= \Delta v_E \cos\theta\, \hat{\theta} + \Delta v_E \sin\theta\, \hat{r} \\ \vec{v}_i &= (v_E + \Delta v_E \cos\theta)\hat{\theta} \\ &\qquad + \Delta v_E \sin\theta\, \hat{r} \end{aligned} \tag{22}$$

Where

$$\begin{aligned} R_E &= 1.500\ x\ 10^{11}\ m \\ v_E &= 30{,}000\ m/s \end{aligned} \tag{23}$$

We optimize $\theta$ to minimize the time of flight to Mars. When the rocket reaches the orbit of mars, we cancel the radial velocity and adjust the angular velocity to be that of mars' orbital velocity.

$$\begin{aligned} \vec{r}_f &= R_M \hat{r} \\ \overrightarrow{\Delta v_M} &= (v_M - \Delta v_M \cos\theta)\hat{\theta} \\ &\qquad + \Delta v_M \sin\theta\, \hat{r} \end{aligned} \tag{24}$$

Where

$$\begin{aligned} R_M &= 2.286\ x\ 10^{11} m \\ v_M &= 24{,}000\ m/s \end{aligned} \tag{25}$$

| **Table 10**: Time of Flight to Mars | | | | |
|---|---|---|---|---|
| Hydrogen, $A$=10m$^2$, $\alpha$=0.9<br>ISP=1466s, $v_{ex}$=14366m/s | | | | |
| Mass Ratio | Time (days) | Angle | $\beta$ | $\frac{\Delta M_E}{M_{prop}}$ |
| 1.00 | 110.4 | 65.7 | 0.3470 | 0.428 |
| 2.00 | 62.2 | 83.5 | 0.1663 | 0.250 |
| 4.00 | 40.3 | 87.8 | 0.0783 | 0.148 |
| 8.00 | 29.0 | 89.5 | 0.0420 | 0.099 |
| Hydrogen, $A$=100m$^2$, $\alpha$=0.9<br>ISP=1870s, $v_{ex}$=18324m/s | | | | |
| 1.00 | 81.8 | 75.0 | 0.2480 | 0.316 |
| 2.00 | 41.4 | 85.0 | 0.1090 | 0.169 |
| 4.00 | 31.3 | 88.0 | 0.0550 | 0.106 |
| 8.00 | 22.7 | 89.0 | 0.0250 | 0.060 |
| Hydrogen, $A$=1000m$^2$, $\alpha$=0.9<br>ISP=2282s, $v_{ex}$=22367m/s | | | | |
| 1.00 | 64.7 | 82.7 | 0.1783 | 0.233 |
| 2.00 | 38.4 | 88.3 | 0.0715 | 0.113 |
| 4.00 | 25.7 | 89.5 | 0.0330 | 0.065 |
| 8.00 | 18.7 | 89.8 | 0.0178 | 0.043 |
| | | | | |
| Water, $A$=100m$^2$, $\alpha$=0.9<br>ISP=856s, $v_{ex}$=8389m/s | | | | |
| 2.00 | 122.9 | 61.6 | 0.3829 | 0.515 |
| 4.00 | 75.8 | 79.3 | 0.2257 | 0.381 |
| 8.00 | 52.3 | 85.7 | 0.1253 | 0.271 |

Where we have the constraint that the total change in velocity is:

$$\Delta v_E + \Delta v_M = \Delta v_{ex} = v_{ex} ln\left(1 + \frac{M_{prop}}{M_{rocket}}\right) \quad (26)$$

And

$$\Delta v_E = \beta \Delta v_{ex}$$
$$\Delta v_M = (1-\beta)\Delta v_{ex} \quad (27)$$

where $\beta$ is adjusted to meet the condition above, and the initial Earth's orbital boost propellant mass, denoted $\Delta M_E$, is computed from,

$$\frac{\Delta M_E}{M_{prop}} = \frac{(1+\Gamma)}{\Gamma}\left[1-(1+\Gamma)^{-\beta}\right] \quad (28)$$

where $\Gamma = M_{prop}/M_{rocket}$ . In Table 10 we show our results for different propellants and fuels to rocket mass ratios.

## VII. Discussion

We have demonstrated a methodology to significantly increase the efficiency of a nuclear thermal rocket through the addition of an electromagnetic accelerator. However, there are several challenges that need to be resolved in the development of a successful NTE thruster. The first challenge will be the development of a suitable electromagnetic accelerator. This accelerator will have to possess three key features:

1. It must be able to handle high variable gas flow rates.
2. It must be capable of handling substantial power levels.
3. It must be efficient.

Two electromagnetic thrusters that meet these requirements are the VASMIR engine, References [9, 10], and the MHDP Thruster, Reference [7, 8]. However, it is important to note that achieving higher power and efficiency would necessitate significant development, posing technical challenges that are nevertheless feasible

A second challenge in the development of the NTE thruster involves creating a high-temperature turbine capable of withstanding the flow rate and temperatures associated with this system. Notably, progress in developing high-temperature turbines for jet aircraft [15-17] has already been made, providing a strong foundation for expediting their adaptation to this purpose. The third obstacle revolves around the development of high-temperature radiators [18]. Fortunately, this is expected to pose minimal difficulty as the input temperature, based on simulations, is at most 2000K, allowing for the utilization of several suitable materials. Finally, the integration of the entire system into a compact engine suitable for interplanetary travel remains a critical aspect of the project

While it is important to acknowledge that this investigation is at a very preliminary stage, we are optimistic about its potential to offer a valuable framework for subsequent investigations and technological advancements

## VIII. Acknowledgements

I would like to Dr. Clancy Weeks for helpful discussions and help in preparing this manuscript.

# Tables

| **Table 1**: Performance Results for $A=10m^2$ and $P_3=1.8bar$ Standard Reference ISP=975s | | | | | |
|---|---|---|---|---|---|
| $\alpha$ | Electrical Power (MW) | $P_2$ (bar) | Velocity (m/s) | ISP (s) | Thrust (kN) |
| 0.000 | 36.75 | - | 12114 | 1236 | 12.11 |
| 0.100 | 33.92 | 23.9 | 12207 | 1246 | 10.99 |
| 0.200 | 30.78 | 32.5 | 12271 | 1252 | 9.816 |
| 0.300 | 27.48 | 38.3 | 12335 | 1259 | 8.635 |
| 0.400 | 24.10 | 42.8 | 12409 | 1266 | 7.445 |
| 0.500 | 20.67 | 46.3 | 12503 | 1276 | 6.251 |
| 0.600 | 17.20 | 49.3 | 12634 | 1289 | 5.054 |
| 0.700 | 13.70 | 51.7 | 12844 | 1311 | 3.853 |
| 0.800 | 10.18 | 53.9 | 13244 | 1351 | 2.648 |
| 0.900 | 6.637 | 55.8 | 14366 | 1466 | 1.437 |
| 0.950 | 4.863 | 56.6 | 16374 | 1671 | 0.819 |
| 0.975 | 3.974 | 57.0 | 19787 | 2019 | 0.495 |

| **Table 2**: Performance Results for $A=100m^2$ and $P_3=1.8bar$ Standard Reference ISP=975s | | | | | |
|---|---|---|---|---|---|
| $\alpha$ | Electrical Power (MW) | $P_2$ (bar) | Velocity (m/s) | ISP (s) | Thrust (kN) |
| 0.000 | 36.75 | - | 12114 | 1236 | 12.11 |
| 0.100 | 34.79 | 8.20 | 12286 | 1253 | 11.06 |
| 0.200 | 32.50 | 10.8 | 12445 | 1270 | 9.956 |
| 0.300 | 30.00 | 13.3 | 12623 | 1288 | 8.836 |
| 0.400 | 27.37 | 15.3 | 12840 | 1310 | 7.704 |
| 0.500 | 24.64 | 17.1 | 13123 | 1339 | 6.562 |
| 0.600 | 21.84 | 18.7 | 13522 | 1380 | 5.409 |
| 0.700 | 18.98 | 20.2 | 14148 | 1444 | 4.244 |
| 0.800 | 16.07 | 21.5 | 15307 | 1562 | 3.061 |
| 0.900 | 13.11 | 22.7 | 18324 | 1870 | 1.832 |
| 0.950 | 11.61 | 23.3 | 23198 | 2367 | 1.160 |
| 0.975 | 10.86 | 23.6 | 30703 | 3133 | 0.767 |

| **Table 3**: Performance Results for $A=1000m^2$ and $P_3=1.8bar$<br>Standard Reference ISP=975s | | | | | |
|---|---|---|---|---|---|
| $\alpha$ | Electrical Power (MW) | $P_2$ (bar) | Velocity (m/s) | ISP (s) | Thrust (kN) |
| 0.000 | 36.75 | - | 12114 | 1236 | 12.11 |
| 0.100 | 35.58 | 2.20 | 12357 | 1261 | 11.12 |
| 0.200 | 34.15 | 3.20 | 12611 | 1287 | 10.08 |
| 0.300 | 32.55 | 3.90 | 12909 | 1317 | 9.035 |
| 0.400 | 30.81 | 4.60 | 13280 | 1355 | 7.970 |
| 0.500 | 28.98 | 5.20 | 13768 | 1405 | 6.884 |
| 0.600 | 27.07 | 5.70 | 14457 | 1475 | 5.783 |
| 0.700 | 25.10 | 6.20 | 15523 | 1584 | 4.647 |
| 0.800 | 23.07 | 6.60 | 17447 | 1780 | 3.489 |
| 0.900 | 21.00 | 7.10 | 22215 | 2266 | 2.222 |
| 0.950 | 19.94 | 7.30 | 29515 | 3012 | 1.476 |
| 0.975 | 19.41 | 7.40 | 40326 | 4115 | 1.008 |

| **Table 4**: Performance Results for $A=10m^2$ and $P_3=1.8bar$,<br>Standard Reference ISP=565s | | | | | |
|---|---|---|---|---|---|
| $\alpha$ | Electrical Power (MW) | $P_2$ (bar) | Velocity (m/s) | ISP (s) | Thrust (kN) |
| 0.000 | 13.57 | - | 7376 | 753 | 7.376 |
| 0.100 | 12.66 | 41.1 | 7442 | 758 | 6.700 |
| 0.200 | 11.58 | 52.3 | 7497 | 765 | 6.000 |
| 0.300 | 10.44 | 59.8 | 7556 | 771 | 5.289 |
| 0.400 | 9.269 | 65.5 | 7626 | 778 | 4.576 |
| 0.500 | 8.071 | 70.1 | 7717 | 787 | 3.858 |
| 0.600 | 6.855 | 74.0 | 7845 | 800 | 3.138 |
| 0.700 | 6.624 | 77.3 | 8047 | 821 | 2.414 |
| 0.800 | 4.381 | 80.3 | 8430 | 860 | 1.686 |
| 0.900 | 3.128 | 82.9 | 9478 | 967 | 0.948 |
| 0.950 | 2.500 | 84.1 | 11279 | 1151 | 0.564 |
| 0.975 | 2.183 | 84.7 | 14210 | 1450 | 0.355 |

| **Table 5**: Performance Results for $A=100m^2$ and $P_3=1.8bar$, Standard Reference ISP=565s | | | | | |
|---|---|---|---|---|---|
| $\alpha$ | Electrical Power (MW) | $P_2$ (bar) | Velocity (m/s) | ISP (s) | Thrust (kN) |
| 0.000 | 13.57 | - | 7376 | 753 | 7.376 |
| 0.100 | 12.93 | 17.2 | 7485 | 764 | 6.735 |
| 0.200 | 12.13 | 22.5 | 7589 | 774 | 6.071 |
| 0.300 | 11.27 | 26.3 | 7712 | 787 | 5.400 |
| 0.400 | 10.37 | 29.3 | 7863 | 802 | 4.718 |
| 0.500 | 9.435 | 31.9 | 8062 | 823 | 4.031 |
| 0.600 | 8.476 | 34.1 | 8345 | 852 | 3.338 |
| 0.700 | 7.495 | 36.1 | 8788 | 897 | 2.636 |
| 0.800 | 6.500 | 37.9 | 9604 | 980 | 1.921 |
| 0.900 | 5.485 | 39.6 | 11703 | 1194 | 1.170 |
| 0.950 | 4.973 | 40.3 | 15039 | 1535 | 0.752 |
| 0.975 | 4.716 | 40.7 | 20113 | 2052 | 0.503 |

| **Table 6**: Performance Results for $A=1000m^2$ and $P_3=1.8bar$, Standard Reference ISP=565s | | | | | |
|---|---|---|---|---|---|
| $\alpha$ | Electrical Power (MW) | $P_2$ (bar) | Velocity (m/s) | ISP (s) | Thrust (kN) |
| 0.000 | 13.57 | - | 7376 | 753 | 7.376 |
| 0.100 | 13.16 | 6.90 | 7517 | 767 | 6.765 |
| 0.200 | 12.63 | 9.10 | 7670 | 783 | 6.136 |
| 0.300 | 12.04 | 10.7 | 7851 | 801 | 5.496 |
| 0.400 | 11.41 | 12.0 | 8080 | 824 | 4.847 |
| 0.500 | 10.75 | 13.1 | 8382 | 855 | 4.191 |
| 0.600 | 10.07 | 14.1 | 8810 | 899 | 3.524 |
| 0.700 | 9.373 | 15.0 | 9474 | 967 | 2.842 |
| 0.800 | 8.658 | 15.8 | 10670 | 1089 | 2.134 |
| 0.900 | 7.928 | 16.5 | 13631 | 1391 | 1.363 |
| 0.950 | 7.557 | 16.9 | 18154 | 1852 | 0.907 |
| 0.975 | 7.371 | 17.1 | 24838 | 2535 | 0.621 |

| **Table 7**: Performance Results for $A=10m^2$ and $P_3=1.8bar$, Standard Reference ISP=375s | | | | | |
|---|---|---|---|---|---|
| $\alpha$ | Electrical Power (MW) | $P_2$ (bar) | Velocity (m/s) | ISP (s) | Thrust (kN) |
| 0.000 | 4.881 | - | 4423 | 451 | 4.423 |
| 0.100 | 4.657 | 2.2 | 4492 | 458 | 4.043 |
| 0.200 | 4.364 | 3.5 | 4554 | 465 | 3.643 |
| 0.300 | 4.038 | 4.6 | 4622 | 472 | 3.235 |
| 0.400 | 3.691 | 5.5 | 4704 | 480 | 2.823 |
| 0.500 | 3.329 | 6.3 | 4811 | 491 | 2.405 |
| 0.600 | 2.955 | 7.1 | 4960 | 506 | 1.984 |
| 0.700 | 2.571 | 7.8 | 5193 | 530 | 1.558 |
| 0.800 | 2.180 | 8.5 | 5624 | 574 | 1.125 |
| 0.900 | 1.781 | 9.2 | 6742 | 688 | 0.674 |
| 0.950 | 1.580 | 9.5 | 8546 | 872 | 0.427 |
| 0.975 | 1.478 | 9.6 | 11319 | 1155 | 0.283 |

| **Table 8**: Performance Results for $A=100m^2$ and $P_3=1.8bar$, Standard Reference ISP=375s | | | | | |
|---|---|---|---|---|---|
| $\alpha$ | Electrical Power (MW) | $P_2$ (bar) | Velocity (m/s) | ISP (s) | Thrust (kN) |
| 0.000 | 4.881 | - | 4423 | 451 | 4.423 |
| 0.100 | 4.769 | 1.8 | 4519 | 461 | 4.067 |
| 0.200 | 4.584 | 1.8 | 4614 | 471 | 3.691 |
| 0.300 | 4.398 | 1.8 | 4732 | 483 | 3.312 |
| 0.400 | 4.211 | 1.8 | 4885 | 498 | 2.931 |
| 0.500 | 4.009 | 1.8 | 5085 | 519 | 2.543 |
| 0.600 | 3.779 | 1.8 | 5359 | 547 | 2.144 |
| 0.700 | 3.538 | 1.8 | 5780 | 590 | 1.734 |
| 0.800 | 3.287 | 1.9 | 6534 | 667 | 1.307 |
| 0.900 | 3.028 | 2.1 | 8389 | 856 | 0.839 |
| 0.950 | 2.895 | 2.2 | 11209 | 1143 | 0.560 |
| 0.975 | 2.828 | 2.2 | 15367 | 1568 | 0.384 |

| **Table 9**: Performance Results for $A=1000m^2$ and $P_3=1bar$, Standard Reference ISP=375s | | | | | |
|---|---|---|---|---|---|
| $\alpha$ | Electrical Power (MW) | $P_2$ (bar) | Velocity (m/s) | ISP (s) | Thrust (kN) |
| 0.000 | 4.881 | - | 4423 | 451 | 4.423 |
| 0.100 | 4.868 | 1.8 | 4544 | 464 | 4.089 |
| 0.200 | 4.868 | 1.8 | 4690 | 479 | 3.752 |
| 0.300 | 4.868 | 1.8 | 4872 | 497 | 3.410 |
| 0.400 | 4.868 | 1.8 | 5104 | 521 | 3.063 |
| 0.500 | 4.764 | 1.8 | 5374 | 548 | 2.687 |
| 0.600 | 4.593 | 1.8 | 5726 | 584 | 2.291 |
| 0.700 | 4.417 | 1.8 | 6267 | 639 | 1.880 |
| 0.800 | 4.239 | 1.8 | 7226 | 737 | 1.445 |
| 0.900 | 4.058 | 1.8 | 9539 | 973 | 0.954 |
| 0.950 | 3.968 | 1.8 | 12982 | 1324 | 0.649 |
| 0.975 | 3.922 | 1.8 | 17989 | 1835 | 0.449 |

# Appendix A: Mathematica Code

Here we show the Mathematic [19] code that we used to calculate the NTE thruster performance

```
(*--Start--*)
(*--Constants--*)
gg = 9.800000000000000000000000000;
RR = 8.3144626180000000000000000;
(*--Specific Heats--*)
Prop="Water";
Which[Prop=="Hydrogen",
    (*--Hydrogen--*)
    AtMss    = 2.01600000000000000000000;
    HeatV   = (1000/AtMss)*0.904000000000000000;
    SHeat   = (1000/AtMss)*0.028000000000000000;
    TVap     = 20;
    TSol     = 14;
    T0       = 100;
    ff1[x_]  = 43.546300000000000000000000/(35+x^(7/10));
    ff2[x_]  = 100+1000*(1-Exp[-0.000075000000000000000000000000*x]);
    ff3[x_]  = ff1[x]*Cos[1.04000000000000000000000*x/ff2[x]];
    CPx[x_]  = (2000/AtMss)*(13.276853691544976854+0.002087896661978l924238*x-1.3573647247785592372*10^(-7)*x^2-ff3[x]);
    CVx[x_]  = CPx[x]-(1000/2)*RR;
    CPx4[x_] = CPx[x]/(x*x*x*x);
    kk[x_]   = CPx[x]/CVx[x];
    (*----*)
,Prop=="Helium",
    (*--Helium--*)
    AtMss    = 4.0025000000000000000000000000;
    HeatV    = (1000/AtMss)*0.0829;
    SHeat    = (1000/AtMss)*0.02078;
    TVap     = 4;
    TSol     = 0;
    T0       = 100;
    CPx[x]   = 5193;
    CVx[x]   = CPx[x]-(1000/AtMss)*RR;
   CPx4[x_] = CPx[x]/(x*x*x*x);
    kk[x_]   = CPx[x]/CVx[x];
  (*----*)
,Prop=="Water",
    (*--Water--*)
    AtMss   = 18.015300000000000000000000000000;
    HeatV   = 2264705;
    SHeat   = 4186;
    TVap    = 373;
    TSol    = 273;
    T0      = 373;
    ff[x_ ] = 33.45707178628649 - 0.004258021235971354*x +  0.0000218297660380414*x^2 - 1.2175728573238403*10^(-8)*x^3 \
        + 2.4063912254431133*10^(-12)*x^4 + 1.0938477204443296*10^(-16)*x^5 - 1.2010607050069152*^-19*(x)^6 \
                                   + 1.7896619728322497*10^(-23)*x^7 - 8.804346818636408*10^(-28)*x^8;
    CPx[x_] = (1000/AtMss)*ff[x];
    CVx[x_] = CPx[x]-(1000/AtMss)*RR;
    CPx4[x_]= CPx[x]/(x*x*x*x);
    kk[x_]  = ff[x]/(ff[x]-RR);
    (*---*)
]
(*--Functions--*)
Adiabatic[Pin_,Pfin_,Tin_] :=
  Module[{PP0=Pin,PP1=Pfin,TT0=Tin},
    NN = 4000;
                 delP = (PP1-PP0)/NN;
                 Told = TT0;
                 Pold = PP0;
                 Do[
                    Pnew = Pold + delP;
                    Tnew = Told*((Pnew/Pold)^((kk[Told]-1)/kk[Told]));
                    Told = Tnew;
                    Pold = Pnew;
```

```
                  ,{i,1,NN}];
                  Tnew
  ]
Radiator[Rin_,AArea_,Tin_,TVap_,LHeat_] :=
  Module[{RR=Rin,AA=AArea,TT1=Tin,TTV=TVap,LH=LHeat},
    Sigma = 5.670374000000000000000*10^(-8);
    kaa   = Sigma*AA/RR;
    aamax = NIntegrate[CPx4[x],{x,TTV,TT1}];
                  LHfac = N[LH/(TTV^4)];
                  If[aamax<kaa,
                     Tnew = TTV-Min[(kaa-aamax)/LHfac,1];
                  ,
                     TT0  = TTV;
       DelT = (TT0+TT1)/2;
       Do[
         aa = NIntegrate[CPx4[x],{x,TT0,TT1}];
                       If[aa>kaa,
                          Tnew = TT0+DelT;
                       ,
                          Tnew = TT0-DelT;
                       ];
                       TT0 = Tnew;
                       DT = DelT/2;
                       DelT = DT;
       ,{i,1,60}];
                  ];
    Tnew
  ]
(*----*)
AArea = 1000;
Rate  = 1;
(*--*)
T1 = 3000;
P1 = 300;
T6 = 3000;
P3 = 1.800000000000000000000000;
PE = 0.010000000000000000000000;
P2   = 1.7000000000000000000000000000000000;
(*----*)
T3 = Adiabatic[P1,P3,T1];
TE = Adiabatic[P3,PE,T6];
(*----*)
alpha=0.1000000000000000000000000000000000;
Rate1 = Rate*(1-alpha);
Rate2 = Rate*alpha;
(*----*)
Print["T0 = ",T0," T1 = ",T1," T2 = ",T2," T3 = ",T3," T4 = ",T4," T5 = ",T5," T6 = ",T6," TE = ",TE];
(*—-*)
PPower1 = Rate1*(NIntegrate[CPx[x],{x,T0,T6}]-NIntegrate[CPx[x],{x,T0,TE}]);
PPower2 = Rate1*(NIntegrate[CPx[x],{x,T0,T1}]-NIntegrate[CPx[x],{x,T0,T3}]);
PPower3 = Rate2*(NIntegrate[CPx[x],{x,T5,T1}]-NIntegrate[CPx[x],{x,T4,T2}]);
(*----*)
Print["Power1 = ",PPower1," Power2 = ",PPower2," Power3 = ",PPower3," Power = ",(PPower2+PPower3)/1000^2];
velE = Sqrt[2*(PPower1+PPower2+PPower3)/Rate1];
THR  = Rate1*velE;
 ISP  = THR/(Rate1*gg);
 Print["Alpha = ",alpha," Velocity = ",velE," THR = ",THR/1000," ISP = ",ISP];
(*----*)
```

# Appendix B: High Temperature Radiators

This is a quick note on a more efficient thermal radiator that could be incorporated into existing space systems. It is more complex than the systems in use today [20], but could be significantly lighter and more compact. For example, the International Space Station has a thermal radiator system that can eject up to 70kW of waste heat using 170m² of cooling area. The system that we propose here can eject close to the same amount heat with a 10m² (2.2m by 2.2m) radiator. In Figure 1B, we show the schematic of our high-temperature radiator as well as its performance specs. The working gas is helium.

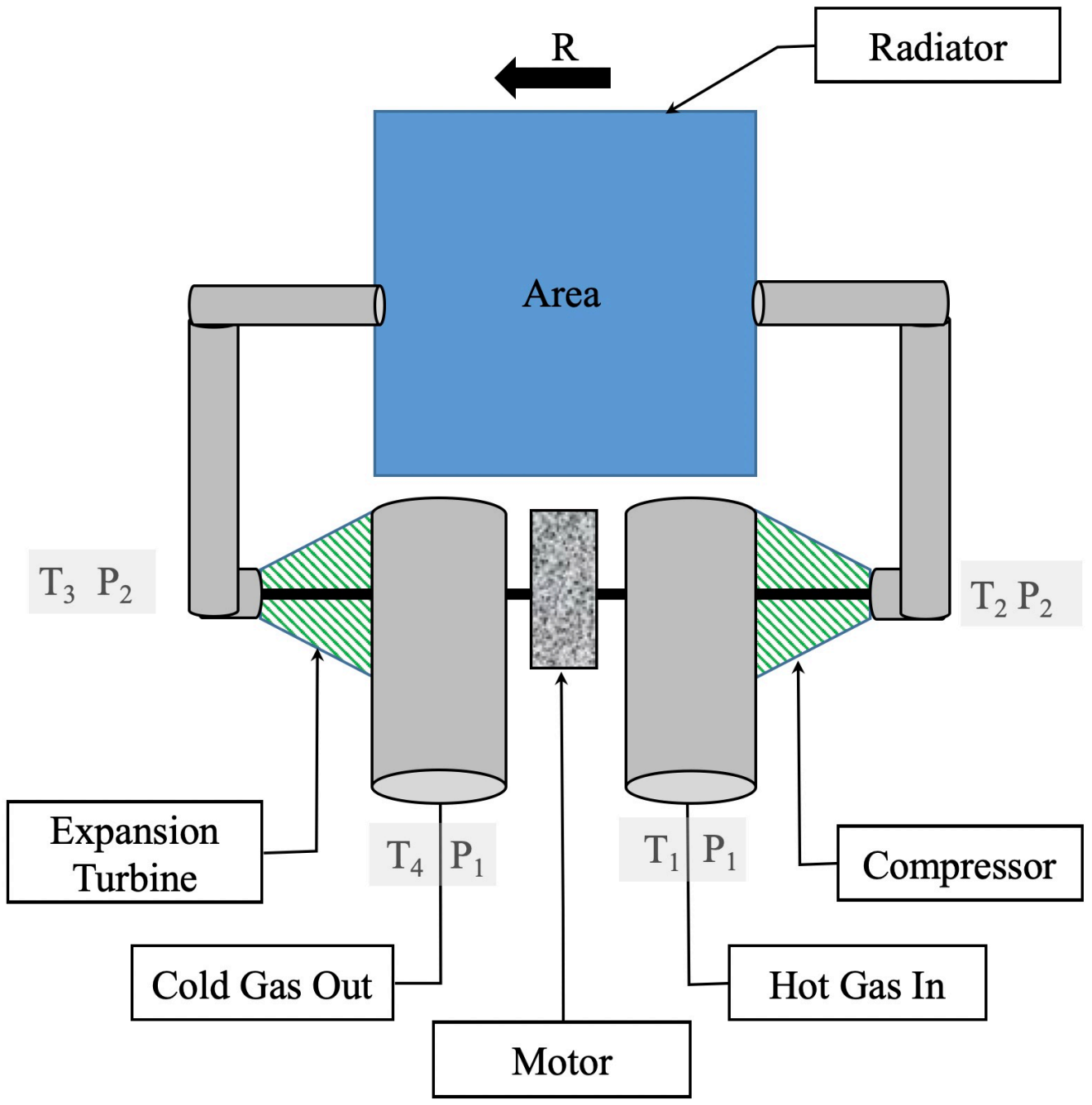

FIG 1B: High temperature radiator schematic

| **Table 1B**: Performance results for the high temperature radiator | | | |
|---|---|---|---|
| *$A=10m^2$, $T_1=300K$, $P_1=1bar$* | | | |
| $P_2/P_1$ | Rate (kg/s) | Cooling Power (kW) | Electrical Power (kW) |
| 2.5 | 0.2 | 12.7 | 5.6 |
| 5.0 | 0.4 | 28.8 | 26.08 |
| 10.0 | 1.0 | 67.3 | 101.0 |
| 20.0 | 2.4 | 154.0 | 357.0 |
| 40.0 | 5.1 | 351.2 | 1186.1 |
| *$A=100m^2$, $T_1=300K$, $P_1=1bar$* | | | |
| $P_2/P_1$ | Rate (kg/s) | Cooling Power (kW) | Electrical Power (kW) |
| 2.5 | 2.0 | 126.9 | 56.34 |
| 5.0 | 4.7 | 292.3 | 264.8 |
| 10.0 | 10.8 | 671.6 | 1016.5 |